# Fully Automatic Liquid Metal Printer towards Personal Electronics Manufacture

**Yi Zheng [1], Zhi-Zhu He [1], Jun Yang [1], Jing Liu [1,2]\***

**1.** Beijing Key Lab of CryoBiomedical Engineering and Key Lab of Cryogenics,

Technical Institute of Physics and Chemistry,

Chinese Academy of Sciences, Beijing, China

**2.** Department of Biomedical Engineering, School of Medicine,

Tsinghua University, Beijing, China

**\*Address for correspondence:**

Dr. Jing Liu

Beijing Key Lab of CryoBiomedical Engineering,

Technical Institute of Physics and Chemistry,

Chinese Academy of Sciences,

Beijing 100190, China

E-mail address: jliu@mail.ipc.ac.cn

Tel. +86-10-82543765

Fax: +86-10-82543767

**Abstract**

Printed electronics is quickly emerging with tremendous value in a wide variety of latest electrical engineering areas. However, restricted to the rather limited conductive inks and printing strategies, the currently existing electronics manufacturing tools are generally complicated, expensive, time, material, water and energy consuming, and thus mainly restricted to the industrial use. Here from an alternative way, the present article demonstrated for the first time an entirely automatic printing system for personal electronics manufacturing through introducing a composite liquid metal ink delivery and printing mechanism to overcome the large surface tension facing the solution, and integrating it with the notebook computer controlling algorithm. With the developed printer, any desired electronically conductive patterns spanning from single wires to various complex structures like integrated circuit (IC), antenna, PCB, RFID, electronic greeting cards, decoration arts, classical buildings (White House, Great Wall etc.) or more do-it-yourself (DIY) circuits were demonstrated to be printed out in a moment with high precision. And the total cost for the whole system has reached personal affordability, which is hard to offer by so far the state of the art technologies. Some fundamental fluid dynamics mechanisms related to the proposed tapping mode enabled reliable printing and adhesion of the liquid metal electronics on the flexible substrate was systematically disclosed through theoretical interpretation and experimental measurements. This clearly "beyond-the-lab" technology and pervasively available liquid metal printer opens the way for large scale home level electronics making in the coming time.

**Keywords:** Printed electronics; Direct writing; Fully automatic liquid metal printer;

Personal electronics; Printed circuit on board; Desktop electronic printer;

Flexible electronics; Consumer electronics; Low cost manufacture; Maker

**Introduction**

To date, the printed electronics has developed flourishingly owing to its superiority of unique flexibility, high production efficiency and low-cost over conventional electronics manufacture technologies. Undoubtedly, the achieved advancements have immensely contributed to the modern electronic industry, especially in those emerging areas like printable transistors [1-3], flexible displays [4], electrode [5,6], sensors [7], antennas [8], radio-frequency identification (RFID) tags [9] and solar cell [10] etc. So far, a variety of functional materials including conductive polymers [11], nanoparticles [12,13], carbon nanotubes [14-16], inorganic semiconductors [17] etc. have been intensively investigated in order to make various desired electronic devices, mainly by way of micro-contact printing [18], screen printing [19], roll-to-roll printing [20] and inkjet printing [21]. Meanwhile, it is also gradually realized that most currently available conductive materials still remain undesirable due to the complicated ink-configuration process, high post-treatment temperature (detrimental to flexible substrates), discontented conductivity (much lower than most metal, especially at room ambience) and unaffordable cost etc. These factors restrain in a large extent the in-depth and ubiquitous adoptions of the existing electronics manufacture methods. In order to break up the bottleneck seriously impeding the large scale practices of the printed electronics, here we present an entire solution to address such urgent need, by foremost developing a tapping mode enabled composite fluid delivery and printing mechanism and incorporating the room-temperature liquid metal as electronic inks to directly and automatically print out various desired electronic targets.

With this liquid metal printer, a series of desired electronically conductive patterns spanning from single wires to various complex structures like integrated circuit (IC), antenna, PCB, RFID, electronic greeting cards, decoration arts, classical buildings (White House, Great Wall etc.) or more do-it-yourself (DIY) circuits and devices were demonstrated to be printed out with high precision in a moment. And the total cost for the machine already reached surprisingly low price. This is almost impossible by so far the most advanced technology since conventional PCB making generally takes long time and is material, water and energy consuming and the existing printed electronics is still far away from the real pervasive direct printing purpose. As the first demonstrated personally based automatic liquid metal printer, this straightforward and evidently "beyond-the-lab" technology opens the way for home level electronics manufacture which is expected to be widely used in the coming daily life.

## Results

### Personal liquid metal printer and its basic working features

The presently set up computer-controlled liquid metal printer (Fig. 1A) is mainly composed of an automatic printing driver, an artifact ink cartridge, two pulleys and a notebook computer loaded with controlling software. The driver with built-in control sensor is capable of duly realizing tapping motion of the cartridge to facilitate delivery of the liquid metal and set aside a printing gap to avoid destroying the pre-patterned electronics. When printing, under software guidance of the selected vector picture, the print driver loading with the developed ink cartridge moves along the X direction and the substrate moves in the Y direction harmonically to accomplish the electronics manufacture. Fig. 1B shows a schematic picture of the printing process and Fig. 1C presents the photograph of the printed RFID made of liquid metal ink with excellent quality. Fig. 1D illustrates a practical device which has combined the currently printed circuit with several necessary IC chips. For making a more complex electronics pattern as desired, the present printer also works rather reliably. What presented in Fig. 1E are electronically conductive structures to represent decorative drawings, or classical building. Clearly, the uniformity, smallest line width etc. of the patterns has already reached a pretty high resolution like 100μm, compared with the most advanced direct writing electronics. Readers can find such printing resolution and the influencing factors in later section. Meanwhile, an appendix with supplementary information for the practical printings of various conductive patterns was also provided as follows for the readers to grasp more specific information of the personal liquid metal printer.

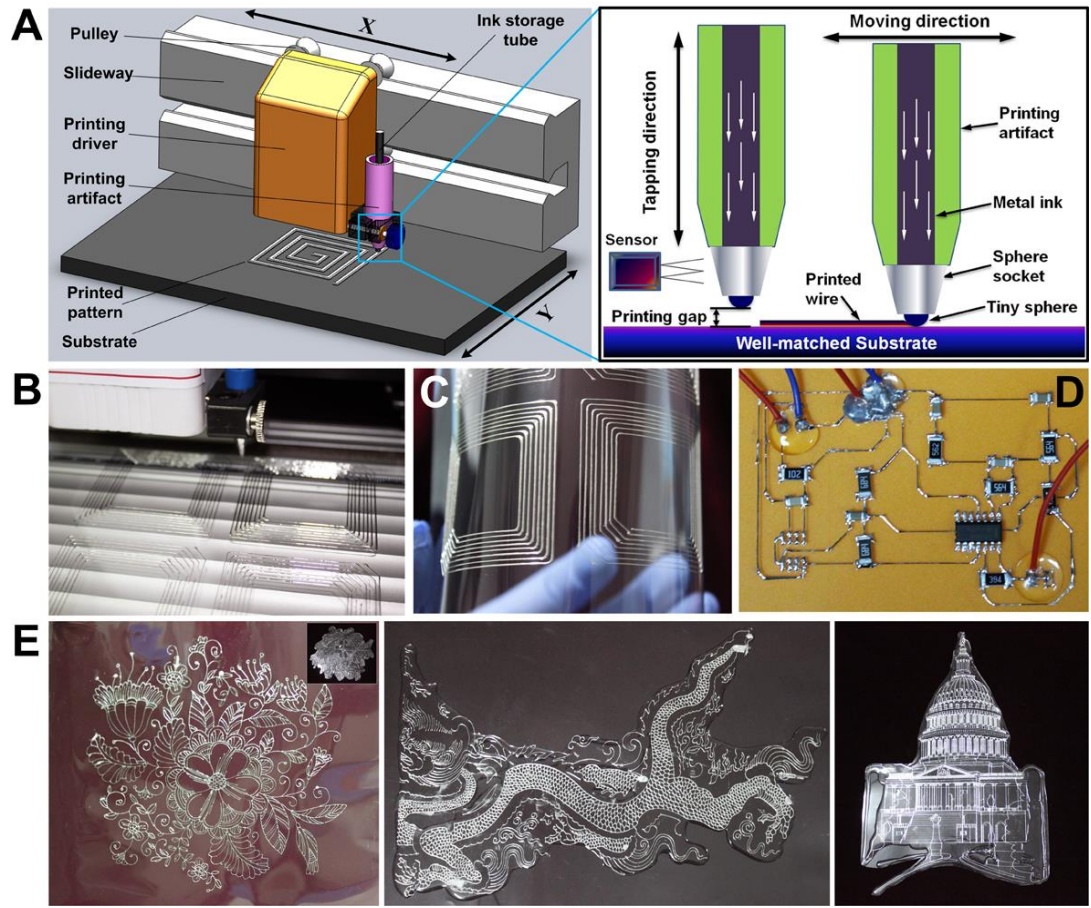

**Figure 1| Schematic of the established liquid metal printer and the printed electronics pattern. A,** 3D diagram of the low-cost computer-controlled printer to illustrate the taping mode movement of the printing head, the composite fluid mechanics to deliver, transfer and adhere the

liquid metal ink to the substrate. **B,** Physical photograph for the printing process of machine. **C,** Photograph of the RFID structure printed by liquid metal ink on flexible PVC film. **D,** Printed circuit with IC chips incorporated; **E,** Directly printed and packaged patterns as decoration drawings and building. (© Jing Liu)

**Composite fluid delivering mechanism to transfer and print liquid metal ink**

The mechanically controlled ink rolling-transfer and printing mechanism is a core to guarantee the reliable running of the current high performance liquid metal printer. The roller-bead (700 μm in size) in the socket (Fig. 2A) is enforced from pressure of the liquid metal ink to tightly fit with the gap to prevent outflow of the ink in the non-operating state (Fig. 2B, C). As printing being started-up through notebook computer, the printing head is driven to move at a specific direction and velocity. Meanwhile, the roller-bead taps to the substrate and the gap is then opened to allow ink outflow with the assistance of roller-bead rotation. Under pressure from the driver, the liquid metal ink with high density are brought out along with the rolling bead and then transferred to the well-matched substrate. The rhythmic tapping action of the roller-bead brings along the oscillation of roller-bead around the gap to remove impurity and maintain smooth flow. The fluid dynamics of the liquid metal in the working state is presented in Fig. 2 (B-E), where the simulated flow streamlines of liquid metal inks under different pressure and speed conditions are illustrated (more detailed analysis is available in our appendix material). To evaluate the impact of the gap width induced by the printing head's tapping mode on the outflow of the ink, additional simulations were also performed. Figure 2F presents the outflow flux of the liquid metal ink through the gap for different driven pressures and the gap widths. Clearly, the gap width plays a critical role in determining the flow flux of the ink. When the gap width tends to zero, the liquid metal ink stops outflow. Overall, the smooth outflow of the ink through the gap is dominated by the driven pressure, the rotation of the roller-bead, and the gap width which all were self-adjusted by the tapping mode. This fundamental core printing mechanism for the liquid metal printer can be attributed to five key steps which were summarized as: 1. Optical sensor to detect the distance between the moving bead and the printing substrate so as to precisely control the tapping mode of the printing head through computer feedback mechanism; 2. Planar mechanical movement of the printing head to guarantee a highly stable rolling of roller-bead during printing and reducing its sliding activity (Fig. 2C, D); 3. Tapping and gravity enabled delivery of liquid metal to the tip bottom (Fig. 2B); 4. Automatically justified smooth flow of liquid metal through the gap between the roller-bead and its surrounding seat (Fig. 2E); and 5. The self-adapted balance between surface tension of the liquid metal and surface energy of the substrate (restricting the stable shape of liquid metal line) guarantees the reliable working of the printing machine. It is also quite noteworthy that the wettability of the liquid metal ink on the printing substrate must be better than the stainless steel printing bead. Only in this way can the inks be smoothly dropped off from the roller-bead surface and adhered to the substrate.

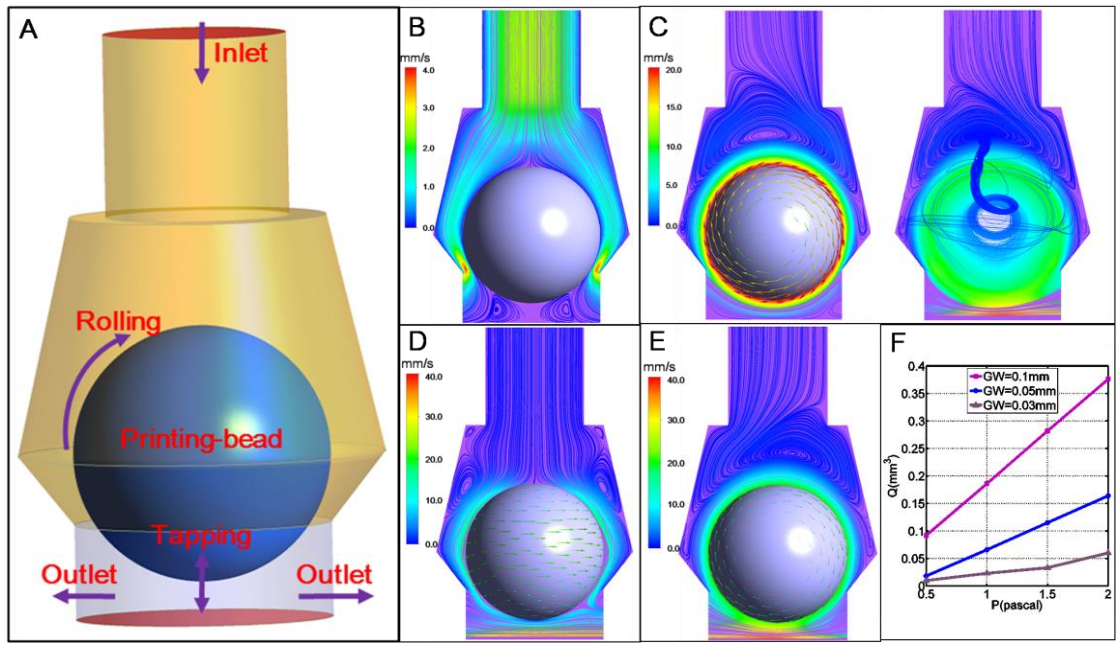

**Figure 2| Composite fluid mechanics for delivering, transferring and printing the liquid metal inks. A,** Illustration for liquid metal delivery through the small gap between roller-bead and its surrounding seat; The flow streamline of the liquid metal ink in the gap for the conditions: **B,** The driven pressure from the inlet is used for overcoming the flow resistance in the socket, $P_{in}-P_{out}=1Pa$, V=0 and $\omega=0$; **C,** The rotation of the roller-bead can enhance the ink outflow through the gap, $P_{in}-P_{out}=1Pa$, V=0 and $\omega=60rad/s$ along the y axis; **D,** When roller-bead rotating along the z axis and completely sliding on the substrate, it has no contribution to the outflow of the ink and induces the printing failure, $P_{in}-P_{out}=1Pa$, V=0.004m/s and $\omega=60rad/s$ along the z axis; **E,** A stable composite fluid transfer state plays a key role in the high-quality printing, $P_{in}-P_{out}=1Pa$, V=0.004m/s and $\omega=60rad/s$ along the y axis. **F,** The outflow flux of the liquid metal ink for different driven pressures (P) and gap widths (GW). (More and detailed explanations on the parameters are available in our appendix materials) (© Jing Liu)

**Wetting properties between liquid metal ink and matched printing substrate**

The wettability of $GaIn_{24.5}$ ink on printing substrate is a key factor affecting the printing quality. Several "paper like" materials were comparatively tested to identify a well-matched printing material. So far, the flexible PVC thin-film exhibits the best outcomes. To further clarify the mechanisms, the contact angles of $GaIn_{24.5}$ ink on universal office paper, "paper like" flexible transparent polyvinyl chloride (PVC) thin film, stainless steel were measured, respectively, to quantitatively evaluate their compatibility with $GaIn_{24.5}$ ink. The sitting drop method was applied to form the liquid metal micro drop and five points fitting method was used to compute all the acquired contact angles. Afterwards, increasing external minute-pressure was clamped down to evaluate the change of the contact angles. Fig. 3A exhibits the measured results of the liquid metal ink drop on the above-mentioned three materials under diverse exerted pressures. It can be obviously discovered that when there is no external force, $GaIn_{24.5}$ ink actually shows apparent non-wetting phenomenon on all surfaces, ($\Theta > 90$ ℃) which is considered to mainly result from the large surface tension of liquid $GaIn_{24.5}$ alloy. Secondly, the applied pressure is able to facilitate wetting of $GaIn_{24.5}$ ink on all substrate materials in varying degrees. Simultaneously, after exerting tiny pressure from upside, the PVC film substrate reveals the best sensitivity to the impressed pressure. With the increasing of the external pressure, the contact angle decreases sharply. However, the office paper just decreases slightly under the same external pressure, implying that the $GaIn_{24.5}$ ink shows much better wetting ability on PVC substrate than office paper, especially under added external pressure. When implementing a pressure of 0.1N, the $GaIn_{24.5}$ ink becomes to exhibit wetting with the PVC film ($\Theta < 90$°). When increasing to 0.2N, $GaIn_{24.5}$ ink reveals excellent compatibility with the PVC film. Therefore, there is a completion described as follows:

Wettability $_{pvc\ film}$ > Wettability $_{steel}$ > Wettability $_{office\ paper}$.

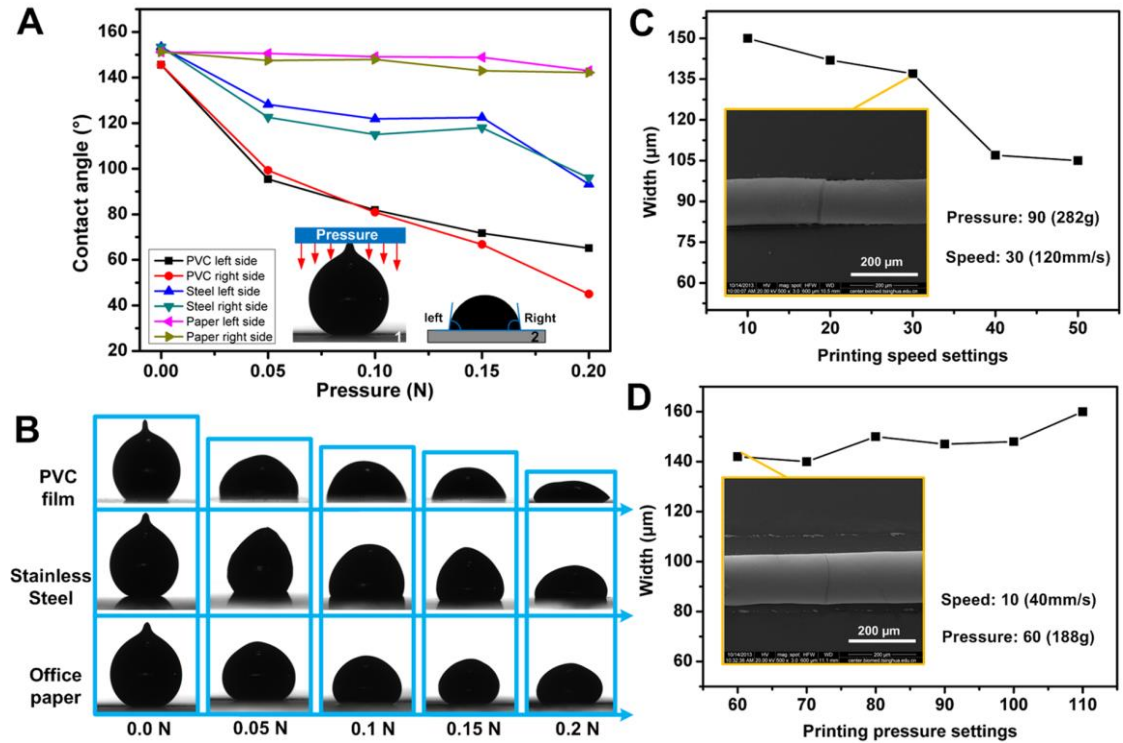

**Figure 3| Effects of external factors to affect the acquired contact angle of GaIn24.5 droplet and the printing resolution. A,** The measured contact angles of GaIn24.5 droplet on PVC plastic film, stainless steel, office paper, respectively, as a function of impressed external minute-pressures increasing from 0.0N to 0.2N. The inset pictures present the schematic of the impressed pressure from upside (1) and illustration of the contact angle in both sides (2) which may be different resulting from a lopsided impressed pressure. **B,** The wetting effect of liquid metal droplets on above-mentioned three kinds of substrate under ascending impressed external pressure, exhibiting that GaIn24.5 droplet reveals excellent wettability on the PVC thin film. **C,** The widths of printed liquid metal track under constant pressure (set as 90°) and increasing printing speed, The inset picture shows the SEM of printed liquid metal circuit at a speed of 30 (120mm/s). **D,** Widths of printed liquid metal track under constant speed (set as 10) and increasing

printing pressure. The inset picture gives the SEM image of printed liquid metal circuit at a speed of 30 (120mm/s). (© Jing Liu)

In accordance with the above-mentioned transfer printing mechanism, the PVC thin film was really a kind of well-matched printing substrate for the presently developed technique. When performing the printing task, the pressure generated from the printing driver subtly precipitates the effluent metal ink to be well printed on the well-matched PVC film.

At this stage, diverse liquid metal flexible electronics have been able to be quickly printed out with high precision through the developed system. This provides an extremely easy going way to make electrical devices which is hard to do otherwise via conventional strategies. The computer-controlled automatic printing allows users to easily make any desired 2D conductive patterns "beyond the lab", just by input the needed images in appropriate vector format. Practically, the printer is able to function in a printing speed of 0~400 mm/s (preferred 0~200 mm/s to ensure the print quality). And a printing pressure in the range of 0~800g was available. Along this way, the resolution of the printed circuits was measured through a scanning electron microscope under various printing speed and percussion pressure, which are both considered to affect the quality. The width values acquired under varied speeds and pressures are presented in Fig. 3C and D, respectively. Fig. 3C shows the widths of printed liquid metal track under constant pressure (set as 90°) and increasing printing speed. Clearly, as the speed increases, the printed line width becomes gradually smaller. The inset picture shows the SEM of printed liquid metal circuit at a speed of 30 (120mm/s). Fig. 3D exhibits the widths of printed liquid metal track under constant speed (set as 10) and increasing printing pressure. There is a slight increase in the width with the ascending of printing pressure. The inset shows the SEM picture of liquid metal line at a pressure of 60 (188g), revealing excellent stability and uniformity.

Meanwhile, it can also be obviously discovered that the slower the speed, the more stable and uniform the lines. When set the printing speed to an over large value, the printed outcome would be completely dissatisfied. Therefore, the printing speed affects more on the quality, available printing speed is preferred to set in the range of 0~50 (0~200mm/s), which can efficiently meet the acquirement in electronic fabrication. High-quality liquid metal line with better resolution (even nanoscale) is currently underway to further promote this process. According to former works, PDMS [25] and room temperature vulcanization (RTV) silicone rubber [28] are both able to act as packaging materials to ensure the environmental and mechanical stability, without affecting the electrical performance. Some packaged electronics can also be found in Fig. 1E and the appendix materials.

**Demonstration experiments on personal electronics manufacture**

So far, the prototype machine of the present liquid metal printer as established in our lab has been able to print out nearly any electrically conductive patterns including spots, lines, curves, and more complex structures. The whole process is entirely automatic and completely controlled by a notebook computer. Basically, the printing of the target circuits depends on its mathematical expressions regarding geometrical domain, which means that the originally printable drawing should be vector graph. Therefore, before printing, the initial electronic draft generally should subject to an evaluation and transformation between different figure formats. Regarding the effect, the currently available high quality printing is already practical enough to fulfill many practical electronics design and manufacture needs.

To demonstrate the diverse capability of the liquid metal printer in manufacturing various electronics drafts, here we choose to print several most typical metal elements which can be used to compose complex electrical circuits or devices. In addition, a few potential areas where personal electronics manufacturing is expected to play important roles were also illustrated through interpreting their potential functions and the fundamental issues lying behind. The order to present the printed conductive targets starts from simple electronic elements to complex patterns until finally functional electronics. As is fully demonstrated, nearly any desired electronically conductive patterns can be directly printed out with high quality through the explored computer-controlled liquid metal printer.

In electrical engineering, a circuit or antenna [29] is generally composed of various basic electrical elements such as straight line, polyline, circle or complex curve. With pre-design and appropriate selection of liquid metal ink, a circuit with desired electrical performances such as size, structure, resistance etc. can be automatically printed out. Figure 4 presented several printed geometrical primitives. Clearly, with a design draft in mind, the corresponding electronic wires or signs (Fig. 4A, C) can be quickly printed out. The present machine also works well for printing characters either in English or Chinese (Fig. 4B). Printed in Fig. 4C are several 2-D conductive patterns. From the perspective of circuit, the printed close structure owns the same potential, which makes the routing manufacture of complex circuit in office rather efficient. Except for making the circuits, the present method also offers the tool for quick production of a series of high-performance electronic sensing elements which is significant for large-scale application of many functional electronic devices. This would provide many attractive applications in the coming time. A quick and straight-forward fabrication of interdigitated-array microelectrodes (IDAM) and flexible antenna by current printer is also demonstrated here, which is very important for making a group of different sensors. An antenna composed of conductive line based on liquid metal ink is presented in Fig. 4D. What one needs here for the printing is only the antenna size such as grid space (4.8mm) and its numbers (7×15) to determine the resonance frequency of the antenna.

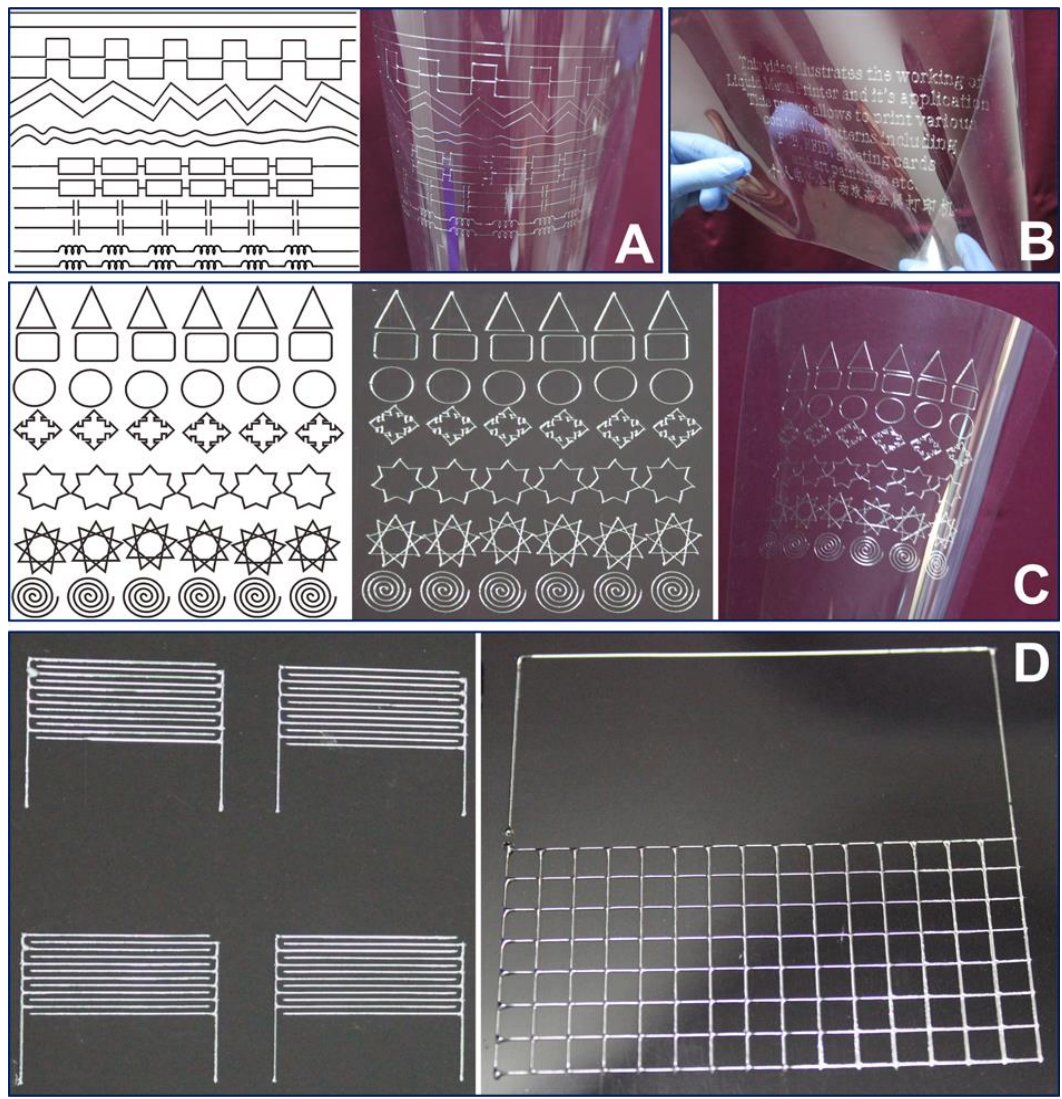

**Figure 4| Printed fundamental elements of lines, curves or characters to compose electronic circuits or devices.** (**A**) Original geometrical design drafts for lines and wires vector graphs and

printed results. (**B**) Printed conductive characters in English and Chinese. (**C**) Pattern vector graph, printed flexible object and its folding state. (**D**) Quick production of high-performance electronic elements such as interdigitated-array microelectrodes or radiofrequency identification antenna.

(© Jing Liu)

Clearly, the present electronics printing method can be applied for manufacturing more other complex electrical devices with additional functions based on the basic concepts enabled herein. In fact, this principle has generalized purpose and can be extended to many different areas even daily life. An expected exciting future would be that people even children without particular training or experiences on electronics can easily print out their own electrical devices in the coming time. The only requirement is just to download electronic drafts, print them out and then assembly all the items together as an integrated device.

Figure 5 exhibits several printed functional patterns made of liquid metal ink. Figure 5A and C show two printed integrated circuits (IC) on transparent PVC film with favorable quality, which took only about 10 minutes. The excellent flexibility of the printed IC is presented in Fig. 5B, which can be adopted for developing electronic devices. This highly efficient electronics fabricating approach consumes no additional energy except for the power to drive the printing cartridge and therefore completely falls in the category of "green manufacture". Fig. 5D exhibits a printed Christmas tree accompanying with the beneath greeting words "Merry Christmas!" which are composed of liquid metal wires illuminated by a few posting-type LEDs, showing the practicability of the printed circuits. A more sophisticated electronics drawing manufacture, electronic painting of the world-famous "Great Wall", is printed on the flexible PVC thin-film and chromatically lighted by LEDs (Fig. 5E). It indicates that the personal electronic devices can now be easily printed out in a moment. The liquid metal line exposed to the air will rapidly form a thin oxide skin to maintain the shape in a degree. The insert picture in Fig. 5E depicts the transient voltages of the two LED lights (1 and 2) with the splashing which shows very stable working property of the printed circuits. Therefore, reliable devices and complicated liquid metal patterns are all able to be manufactured through this straightforward and highly efficient composite printing way. The present method has generalized purpose in making various users oriented electronically conductive patterns. More extensive applications can be found in the appendix materials.

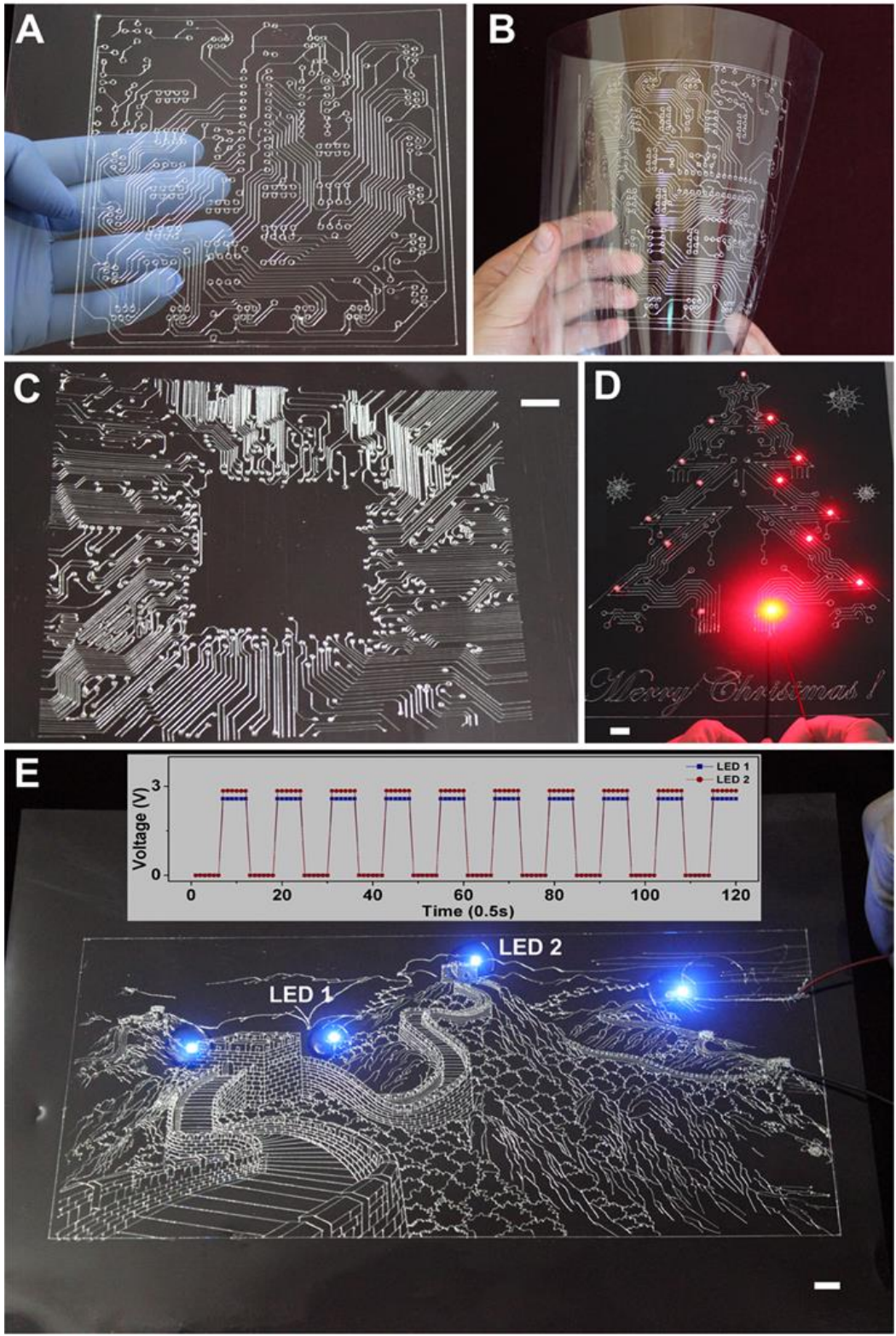

**Figure 5| The quickly printed out functional electronic patterns via liquid alloy printer. A,** Physical picture of a printed-circuits-on-board for IC application. **B**, Folding state of the PCB. **C,** Another different printed integrated circuits on PVC substrate. **D**, Printed "Christmas tree"

composed of liquid metal wires and English characters "Merry Christmas !" is illuminated by several posting-type LEDs. **E,** Manufactured lighting on liquid metal "Great Wall" through the developed method. (All the scale bars are 10 mm) (© Jing Liu)

**Discussion**

As is recently realized, the liquid metal is a perfect electronic ink for directly writing out electrical device. However, a big bottleneck in impeding the successful printing of such solutions lies in its extremely high surface tension. Through a series of comparative experiments, we found that the liquid metal is hard to be delivered inside the cartridge and then deposited on the target substrate via the conventional printing methods such as ink-jet or dispensing printing. As an alternative to those classical principles, which strongly depend on the driven pressure directly applied on the ink, the present liquid metal printing is entirely made possible through the introduction of a composite fluid transfer mechanism which is consisted of a series of complex parallel movement of the printing head.

During the electronics manufacture, the tapping mode movement of the printing head is the core of the method. When the printing-head keeps at stationary state, the static roller-bead is enforced to locate at the bottom of its surround seat due to pressure of the liquid metal, which leads the roller-bead to tightly fit the gap to prevent outflow of the ink. However, the roller-bead rhythmically taps to the substrate during the working state so that the gap is opened continuously to allow ink outflow with the assistance of roller-bead rotation. The rhythmic tapping action induces the oscillation of roller-bead around the gap to remove impurity and maintain flow smooth, which is critical for a long time printing. Compared with the conventional printing method (such as microinject-printing), the tapping mode of the roller-bead owns the function of self-clearance such that it adapts excellently to print the liquid metal ink, which is apt to form metal-oxide (such as gallium oxide) with exposure to air. In addition, the opening amplitude of the gap was adjusted through tapping mode to uniformly control the ink outflow.

In summary, we have demonstrated a pervasive way to automatically print out diverse electronic patterns on transparent plastic film via the invented liquid metal printer. This composite fluid delivery mechanism for printing liquid metal ink is rather convenient, efficient and most important of all, extremely low-priced, implying its huge potentials to be widely adopted in smart electronics fabrication over the world. Its unique virtue is especially important in the increasingly emerging areas of personalized consumer electronics, with the capability of freely making various PCB, integrated circuits (IC), and functional electronics, via the do-it-yourself (DIY) style. In the near future, making little effort and under auxiliary of additional inks made from oxide, nanoparticle, semiconductor and carbon nano-structure, the present composite ink delivery and printing mechanism can naturally be extended to more optoelectronic areas such as printing flexible display, transistors, solar cell, paper batteries and other multifunctional devices. This easily controllable and personally affordable system opens the way for large scale application of the printed electronics and offers a powerful office tool for electronics manufacture in the coming daily life.

**Methods**

**Preparation of liquid metal electronic ink and matching printing substrate**

As a remarkable material with favorable metallic conductivity, flowability, direct printability,

and affordable cost, the EGaIn$_{24.5}$ alloy was introduced here as the electronics ink which owns a melting point of 15.5 ℃ and is therefore capable of remaining in liquid phase under room temperature. Further, its large density (6280 kg/m$^3$) and low viscosity (2.7×10$^{-7}$ m$^2$/s) make GaIn$_{24.5}$ a preferable ink. Such room temperature liquid metals have recently attracted big attentions and were preliminarily utilized to fabricate simple and reliable flexible devices [22-26].

According to the experiments, oxide in the alloy can significantly contribute to the adhesion but immensely reduce the liquidity. To maintain the favorable fluidity of liquid metal alloy, herein, the adopted EGaIn$_{24.5}$ alloy was not deliberately oxidized. For making the electronic ink, high-purity gallium and indium metals (with purity of 99.99 percent) were weighted as source materials with a ratio of 75.5:24.5 in line with the chemical compositions. Then the weighted gallium and indium metals are mixed together in the beaker which was beforehand cleaned by the deionized water and heated to 50 ℃ until metals were fused completely, then stirred slightly. Afterwards, the configured GaIn$_{24.5}$ ink with traces of natural oxide was injected into the ultrasonic cleaned core of the printing head.

Meanwhile, flexible materials suitable as "typical printing substrate" such as commercially available office paper, flexible PVC thin-film were both tried to screen out well-matched printing substrate. A contact angle meter (JC2000D3, Shanghai, China) was used to measure the contact angle of GaIn24.5 on different substrates under different external minute-pressures. Through a comprehensive evaluation, the PVC thin film with both transparent and flexible features was particularly identified as a perfect printing substrate in the present work.

**Set up of the liquid metal printing system**

According to a series of our comparative experiments, the liquid metal inks as fabricated above are extremely hard to be driven by most of the currently available printing technologies such as direct writing, fluid dispensing, and micro-contact printing etc. For example, for the direct-printing ways such as thermal bubble and piezoelectric inkjet printings, the liquid metal ink could not be driven so far, due to its either high vaporization temperature (above 2000 °C) or large surface tension [27, 28]. Therefore from a completely different approach based on composite fluid delivery mechanism (Fig. 1), this study successfully developed an entirely-automatic and cost effective method for printing the liquid metal inks.

Through introducing the tapping mode enabled composite fluid transport mechanism to reliably deliver electronics materials and the printing mechanism of the printing roller-bead, an extremely easy going and low-cost way to achieve fully-automatic printing of room temperature liquid metal ink to make various complicated electronic patterns with high precision was established. Specifically, a printing head pre-loaded with liquid EGaIn$_{24.5}$ alloy was developed. Combining the high density, favorable fluidity, the driving effect of rolling bead and excellent wettability on matched substrate, the present printing head is capable of fluently drawing liquid metal structures on the substrate. Further, to realize a straight forward way for electronics printing just like clicking a mouse to print the required files or photos via printers in the office, a notebook computer controlled plotting platform with innovative printing head, ink and software was set up for the automatic printing of liquid metal. As a result, a diverse range of complex electrically conductive patterns were directly manufactured, in line with the selected drafts stored in the database. At this stage, the total cost for making the machine prototype already reached an extremely low value like hundreds of US dollars. This guarantees the tremendous potential of the technology for large scale personal use.

**Appendix: Supplementary information**

A1. Principles of composite printing process of the liquid metal printer

The current tapping to print mechanism is developed here to tackle the extremely high surface tension of the liquid metal ink (such as 0.624N/m for $GaIn_{24.5}$) [30, 31] and guarantee its uniform and reliable delivery. And our performed series of theoretical and experimental works fully demonstrate the capability of the principle. Basically, the printing of the liquid metal ink is mainly determined by the rolling enabled ink adhesion, transfer and then tapping to press process. Under automatic control of the computer, the printing bead was driven to smoothly roll on the surface of the substrate (Fig. S1A). The liquid metal ink pre-loaded inside the small tubule of the printing head is then uniformly delivered to the tip slot due to gravity effect and adhered to the surface of the roller-bead (Fig. S1B). Subsequently, it is then transferred and finally deposited on the surface of the substrate. The strong force from the upside down tapping motion of the printing-head and the rolling of the printing bead contribute to the extremely tight adhesion of the ink to the target substrate. All these processes are automatically controlled by the computer without any external human interruption. It thus significantly simplifies the difficulty of the electronics manufacture and therefore paved the way for large scale application of the liquid metal printer. The following sections is dedicated to disclose the basic mechanisms of the roller-bead based high quality printing of liquid metal electronics ink to compose various desired circuits.

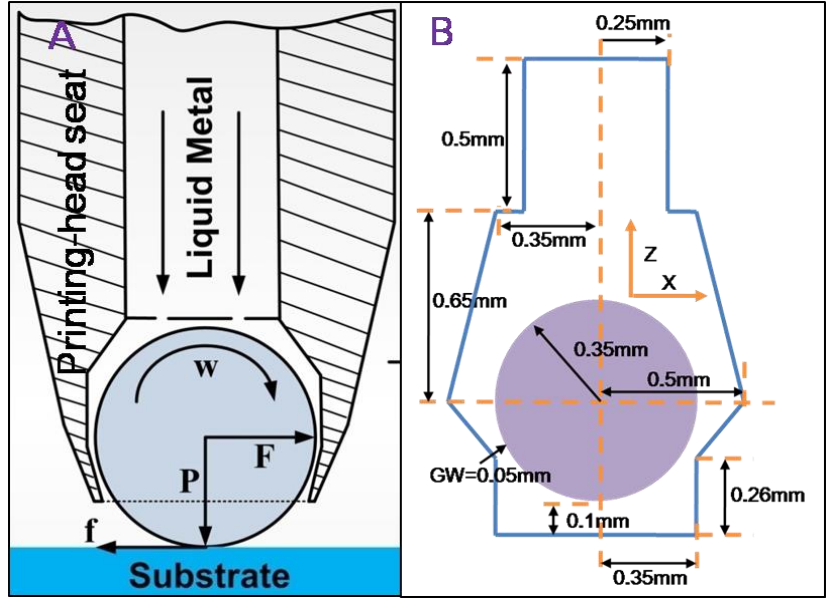

**Figure S1| Schematic diagram of the roller-bead components to deliver and print liquid metal ink.** (**A**) Illustration for liquid metal delivery through the small gap between roller-bead and its surrounding seat; (**B**) Flow domain of the computational model. (© Jing Liu)

A2. Contact mechanics between roller-bead and elastic substrate

During the process for the printing bead to roll on the substrate, liquid metal is continuously transferred and deposited on the substrate surface. Our experiments indicate that a large sliding of the roller-bead would induce flow interruption of the liquid metal through the gap between roller-bead and its seat. Thus the contact mechanics between roller-bead and substrate is very important for guaranteeing the printing quality. Here we investigate to clarify the mechanism of the roller-bead sliding during printing.

Firstly, we consider a static rigid bead contacting with the elastic substrate surface (see insert illustration in Fig. S2) due to suffering from the external force *F*. A local coordinate for the contact surface position can be defined as $z = R^{-1}(x^2 + y^2) - h$, where *R* is the radius of roller-bead, x and y are along the principal curvatures in the tangent plane (denoting here horizontal plane). When roller-bead approaches the substrate surface by a short distance $h$, the deformation displacement $u_z$

of the substrate surface is given by

$$u_z(x, y) = R^{-1}(x^2 + y^2) - h \qquad (S.1)$$

We denote by $P_z$ the pressure at the contact region between the roller-bead and the deformed

substrate. It is obvious that $P_z = 0$ outside the contact region. $P_z$ at the contact region is given by

$$\frac{1-\sigma^2}{\pi E} \int_S \frac{P_z(x, y)}{r} ds = h - R^{-1}(x^2 + y^2) \qquad (S.2)$$

where, $E$ denotes the Young's module and $\sigma$ Poisson's ratios of the substrate. The solution of the

above integral equation can be obtained through analogy with the potential theory [32]. Thus one has

$$P_z(x, y) = \frac{3\cos(\Phi)F}{2\pi a^2} \sqrt{1 - \frac{1}{R}(x^2 + y^2)} \qquad (S.3)$$

where $F$ is the amplitude of the external force, $\Phi$ the angle between the direction of $F$ and z axis, $a$

is the half-width of the indentation (shown in Fig. S2). Furthermore, we can derive the indentation

width of the rigid bead and the radius of the contact region:

$$\begin{cases} h = \left( \dfrac{3E\cos(\Phi)F}{4(1-\sigma^2)\sqrt{R}} \right)^{2/3} \\ a = \sqrt{Rh} \end{cases} \qquad (S.4)$$

From Eq. (S4), it can be found that the contact area satisfies relation $A_c \sim \sqrt{F^3}$ ( $\Phi \approx 0$ is set in our printing experiments), which indicates that force $F$ can be estimated from the contact surface measure. Large contact area $A_c$ can promote the liquid metal to be easily deposited on the substrate when $\gamma_{sub-lm} < \gamma_{sub-sp}$, where $\gamma_{sub-sp}$ is the surface tension of the liquid metal on roller-bead, and $\gamma_{sub-lm}$ is its surface tension on substrate. In fact $\gamma_{sub-lm} < \gamma_{sub-sp}$ has to be enforced so as for the liquid metal to be easily dropped off from the roller-bead surface.

The tangential force for the static roller-bead is given as $\tau = \tan(\Phi)P_z$. No sliding exists for the static roller-bead when $\tau \le \mu P_z$, which means $\tan(\Phi) \le \mu$ ($\mu$ denoting friction coefficient). However, roller-bead experiences an elastic rotation ($\omega$) in addition to the translational movement ($V$) during the printing process. A dimensionless quantity $s = 1 - R\omega/V$ is defined to characterize the sliding of the roller-bead. A detailed derivation of expression$s$ can be found in [33]:

$$s = \frac{-(4-3\sigma)}{4(1-\sigma)} \frac{\mu a}{R} \left[ 1 - \left(1 - \mu^{-1}\tan(\Theta)\right)^{1/3} \right] \tag{S.5}$$

We further define two dimensionless qualities as:

$$\begin{cases} Sr = -4(1-\sigma)sR/(4-3\sigma)\mu a \\ Fr = \mu^{-1}\tan(\Theta) \end{cases} \tag{S.6}$$

Then the relation between $Sr$ and $Fr$ based on Eq.(S5) and (S6) can be found in Fig. S2. It indicated that the external force plays an important key on the sliding of the roller-bead. In addition, the sliding ratio is also dependent on the contact area. From the above analytical interpretation, we have to control the sliding of the roller-bead, which leads to the instable shedding of liquid metal from the roller-bead during printing process. It is noteworthy that increasing the contact area $A_c$ helps liquid metal to well deposit on the substrate, which however also leads to the roller-sphere sliding. Thus, controlling force $F$ is critical for printing the liquid metal ink, which can be fitly adjusted by monitoring the contact area $A_c$ according to Eq. (S4).

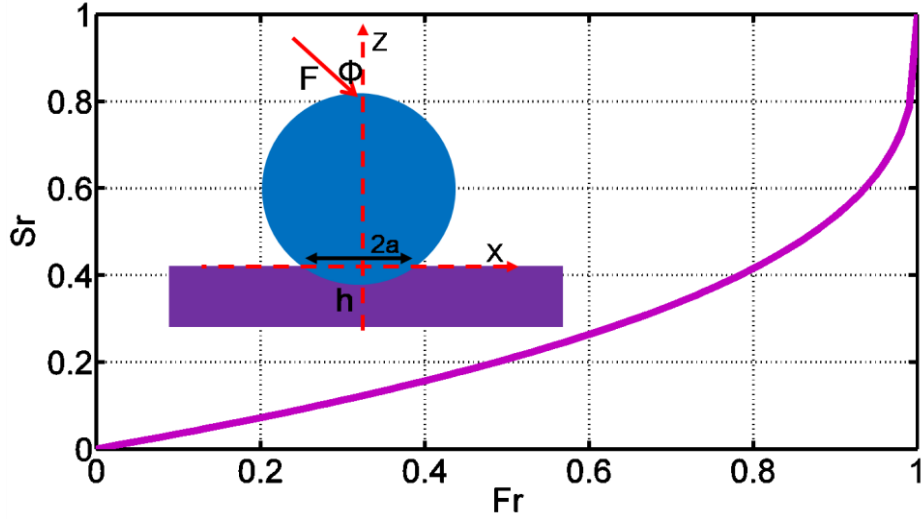

**Figure S2| The relation between *Sr* and *Fr*** (© Jing Liu)

A3. Composite fluid dynamics of liquid metal ink for reliable printing

The outflow of liquid metal through the gap plays a key role in determining the final printing resolution and quality of the printed pattern. It is necessary to investigate the fluid dynamics of liquid metal ink in the gap. A simplified geometrical model of liquid metal flow is depicted as Fig. 1B. The liquid metal ink used here is GaIn$_{24.5}$ (its density: $\rho$=6.28×10$^3$ kg/m$^3$ and viscosity [30, 31]: ν=2.7×10$^{-7}$m$^2$/s) which can be treated as Newton-fluid in small temperature variation case as studied in the present room temperature environment. The governing equation for the liquid metal ink adopts Navier-Stokes equation which reads as:

$$\begin{cases} \nabla \bullet \mathbf{u} = 0 \\ \left( \dfrac{\partial \mathbf{u}}{\partial t} + \mathbf{u} \bullet \nabla \mathbf{u} \right) = -\dfrac{\nabla p}{\rho} + \nu \nabla^2 \mathbf{u} + \mathbf{g} \end{cases} \tag{S.7}$$

where *g* denotes the gravity. The inlet and outlet of computational domains are considered as gauge pressure boundary $P_{in}$=P and $P_{out}$=0, respectively. The reference pressure is set as the standard atmospheric pressure. The rotation velocity on the surface of the roller-bead is assumed as $\mathbf{V_R}$=$\omega$×$R$. In order to characterize the translational motion of the printing head, the relative velocity **V** in x-y plane is given to the substrate surface. The boundary condition on the surface of the roller-bead seat is treated as no-slip. All simulations adopted Fluent 6.3 parallel model and run on the Dell PE2950 workstation with two quad-core CPUs (Intel Xeon x5365 @3.00Hz) and eight memories (1 GB each). In addition, we have performed mesh convergence tests to ensure adequate spatial resolution.

According to the theoretical prediction, the liquid metal ink can flow through the gap due to the pressure difference between inlet and outlet without roller-bead rolling (shown in Fig. S3A, B). When roller-bead rolls on the substrate, the composite flow state (Fig. S3C, D) of the liquid metal is initiated and completely different from that driven by the single pressure. The flow velocity near the outlet of the gap is strongly enhanced by the rotation of the roller-bead, which promotes liquid metal to be dropped off from the roller-bead surface. The fluid flux delivered to the substrate is determined by the flow state and the shortest width between roller-bead and its seat (GW=0.05mm shown in Fig. S1B). In practice, the single pressure cannot drive the liquid metal continuously to flow through the gap due to pretty large surface tension of the liquid metal, which is hard to be considered in the simulation. Benefiting from the rotation of the roller-bead, the liquid metal is transferred rapidly to the substrate and ideally avoids the formation of large contacting surface and thus the high surface tension.

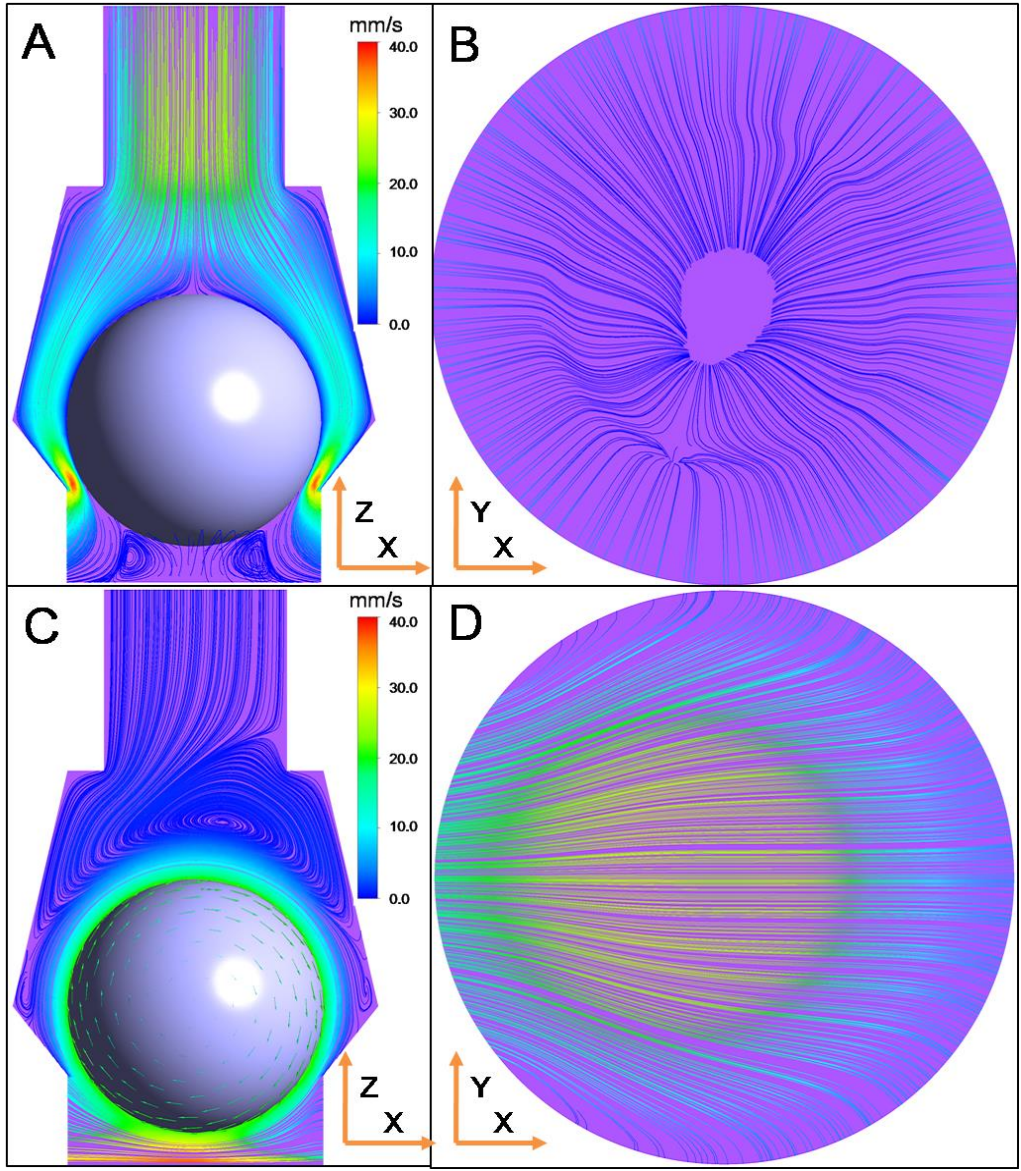

**Figure S3| The flow streamline of the liquid metal ink in the gap**. (**A**) 3D view and (**B**) 2D view at z=-1.5mm cross section with $P_{in}$= 1Pa, $P_{out}$=0, $V$=0 and ω=0; (**C**) 3D view and (**D**) 2D view at z=-1.5mm cross section with $P_{in}$=1, $P_{out}$=0, $V$=40mm/s and ω=70rad/s. (© Jing Liu)

For printing a straight line, the rotation axis of the roller-bead is along y axis, which however may be frequently requested to change such initial direction for printing a nonlinear curve. This often leads to the discontinuous deposition of liquid metal ink to compose target electronic-pattern. Here, an extreme process, roller-bead rotating the z axis and completely sliding on the substrate, is particularly investigated, which may occur during turning around the printing head at the corner. The simulation results are shown in Fig. S4. Clearly, the rotation has no impact on the flow flux of the liquid metal delivering from the gap, which is determined by the pressure. This may make the printing at the turning point failed. Our experimental results also clearly disclosed the flow interruption of the liquid metal during turning at the corner, which was just induced by the rotation failure.

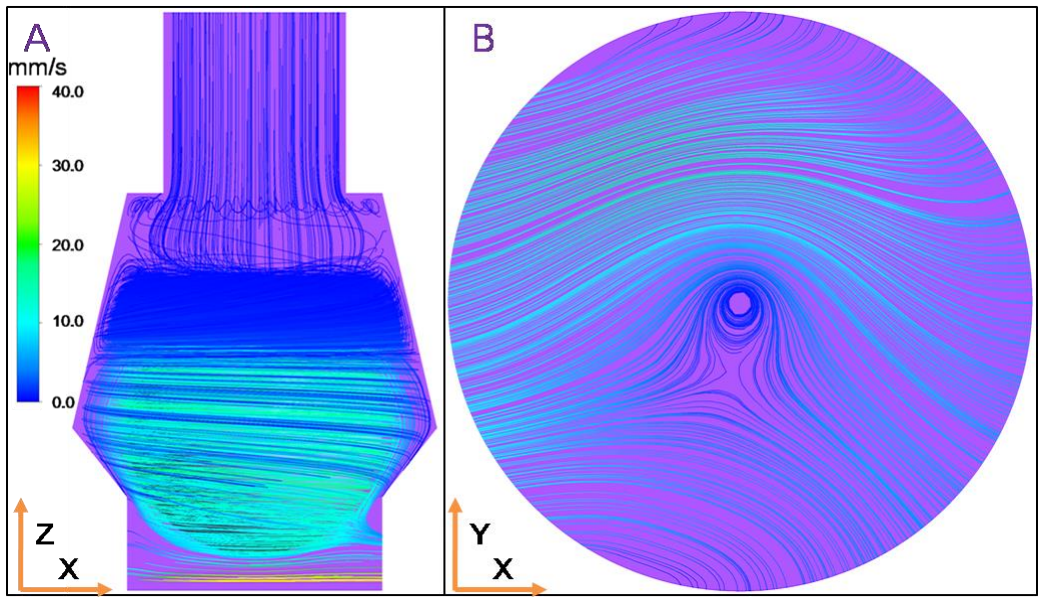

**Figure S4| The flow streamline of the liquid metal ink in the gap**. (**A**) 3D view and (**B**) 2D view at z=-1.5mm cross –section with $P_{in}$= 1Pa, $P_{out}$=0, $V$=40mm/s and $\omega$=60rad/s. (© Jing Liu)

A4. Estimating the stable shape of liquid metal line

After the liquid metal was deposited on the substrate through the roller-bead, the stability of the thin printed pattern is determined by the balance between the surface tension of the liquid metal and the surface energy of the substrate, which is often characterized by the contact angle of the liquid metal droplet on the substrate. The high contact angle (corresponding to less wetting) can lead easily to the instability of the printed pattern [34]. Here, we analyze the matching of liquid metal surface tension with substrate surface energy to investigate the influencing factors of unstable phenomenon of printed pattern. The area of the line conductor cross-section can be approximated as $A=Q/V_s$, where $Q$ is the flow flux of the liquid metal from the gap between roller-bead and its seat. The mechanical equilibrium between liquid metal and substrate can be described by Young's model:

$$\gamma_{sub-air} = \gamma_{sub-lm} + \gamma_{lm-air}\cos(\theta) \tag{S.8}$$

where, $\theta$ is the contact angle of liquid metal with the substrate. According to thermodynamic equilibrium, we can derive the width of stable line conductor $L$, i.e.

$$L = \frac{2\sin(\theta)}{\sqrt{\theta - \sin(\theta)\cos(\theta)}}\sqrt{\frac{Q}{V_s}} \tag{S.9}$$

The above equation indicates that the width of line conductor is mainly determined by the contact angle, flow flux of liquid metal and translational speed of the printer.

For a typical printing process, where V=40mm/s, ω=60rad/s and $P_{in}$-$P_{out}$=1Pa, GW=0.05mm, the flow flux of liquid metal is about Q=0.0656mm3/s (from the above results of fluid dynamics simulation). For θ=40, the width of stable liquid metal line is estimated about L=126μm based on Eq. (S9), which is consistent with the experimental results. Figure S5 presents the relation between the width of stable liquid metal line and the contact angle. It is noteworthy that the printed pattern keeps the liquid state at room temperature, which is easily affected by other factors, such as gravity and vibration. Thus packaging is very important for maintaining a stable electronic pattern for practical purpose.

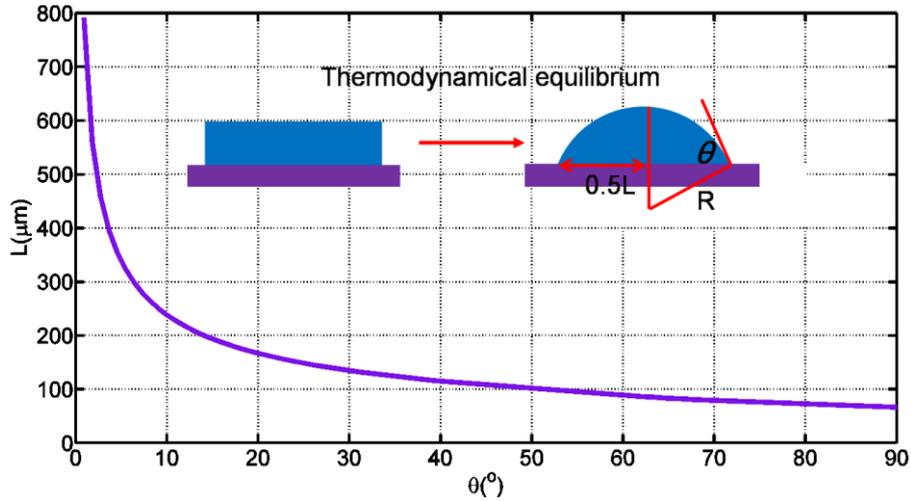

**Figure S5| The relation between the width of stable liquid metal line and contact angle.**

(© Jing Liu)

A5. Diverse capabilities of the liquid metal printer

To demonstrate the wide adaptability of the present composite ink printing method, a series of additional experiments were carried out to manufacture more electronic devices which all exhibit satisfactory outcomes on the well-matched substrate.

The present printed electronics way allows rather wide flexibility. In principle, various complex electrical patterns other than the pure line or curve can be easily printed out. Figure S6 presents several regularly designed 3-dimensonal structures in effect. Here, due to dimension limitation of the current printer, only the projections (Fig. S6(A)) of 3-dimensional objects are printed. In the current manufacture, the printing head moves in X direction, while the substrate moves along Y direction. Therefore X-Y makes up Cartesian coordinate. The printer adopts various composite movements to draw specific target structures. Such printed patterns (Fig. S6(B)) further enrich the routing manufacturing styles.

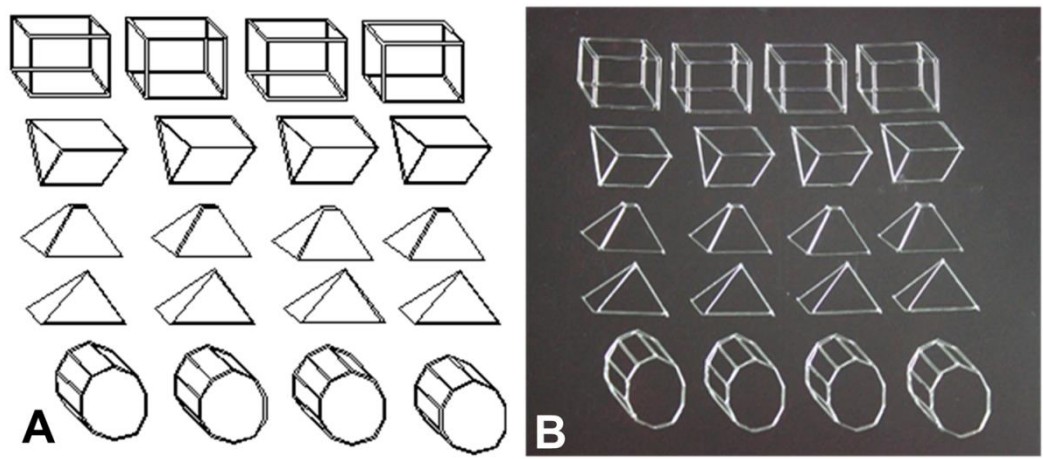

**Figure S6| Regularly designed 3-dimensional structures.** (**A**) Original structure and pattern in vector graph format; (**B**) Printed results according to the draft in (A). (© Jing Liu)

A6. Printing of conductive decoration art

As shown in the above, the present liquid metal printer is capable of manufacturing various electrically conductive lines, curves or patterns which can then be used to compose more complex electrical targets. With such printer in hand, it would help incubate certain emerging consumer electronics. One particular important area may come to the fabrication of electronic decoration drawings. To test the performance of the printer and its potential role for home, office or public use purpose, we choose to print a carefully designed decorative texture as Fig. S7A, B, respectively. Figure S8 presented more printed electronic decoration drawings. Various objects like eagle (Fig. S8A), flower (Fig. S8B) and dragon (Fig. S8C) can be directly printed out in a few minutes. Clearly, with such complex metal pattern, some functional device such as LED as well as IC chip can be incorporated into the circuit. In this way, the IC chip can serve to control the lighting of LED according to the loaded program. Clearly, the combination of printable electronic and artistic field opens exciting new fields for the coming society. For brief, we did not present here the working of such functional electronics but leave the demonstration of the function of the printed electronics to other circuits in later sections.

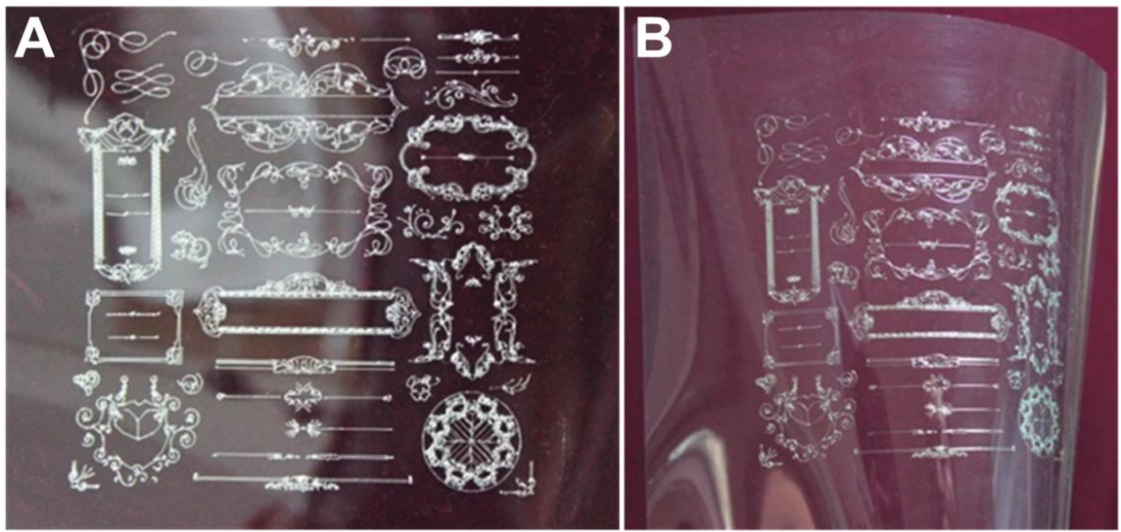

**Figure S7| The electrical decorative textures directly printed by the liquid metal printer.** (**A**) Planar picture; (**B**) Folding state. Such electric drawing is rather appealing for pleasant home decoration since people may wish to design and print their own texture arts themselves. In a certain way, colorful LEDs can be connected with these instantly printed decorative pictures. In dark or weak light circumstance, these arts would look more beautiful when switching on the LEDs light, which makes it terrifically different from the traditional decorative arts. (© Jing Liu)

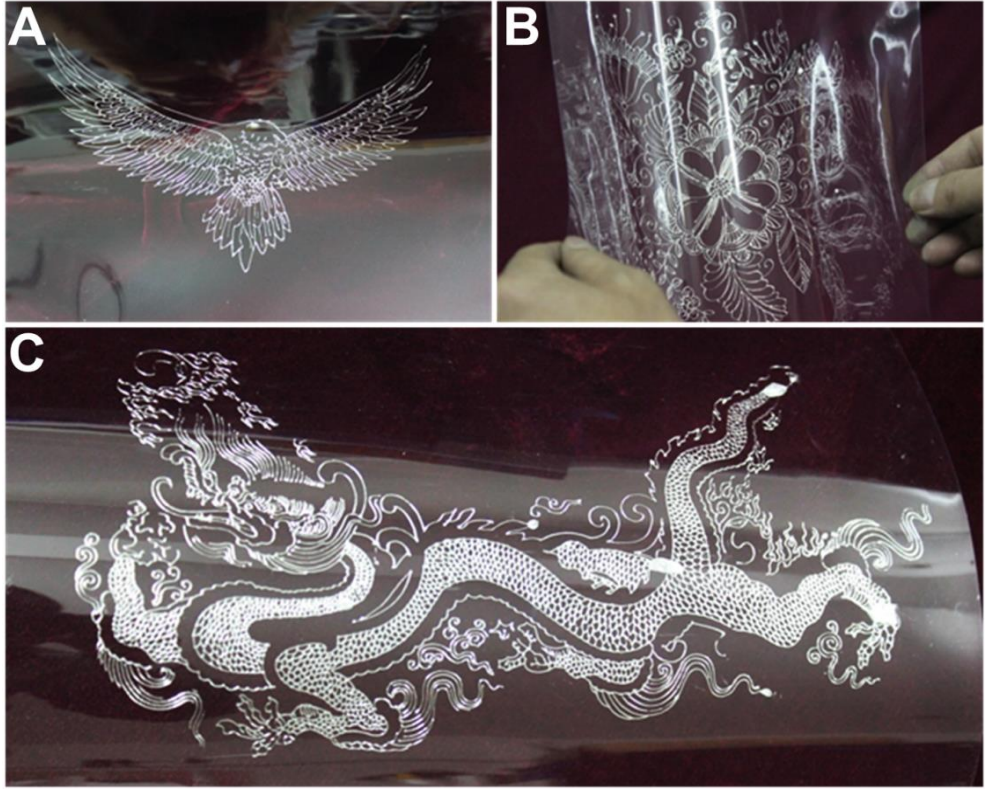

**Figure S8| More printed electronically conductive decoration drawings.** (**A**) Eagle; (**B**) Flower; and (**C**) Dragon. (© Jing Liu)

A7. Printing regularly structured pattern of electrical buildings

To move on, the present machine allows easily print out any desired electrically conductive building pictures. For example, Fig. S9 depicts one of such images of the electrical circuit to duplicate electrically a long historical Chinese building: The Temple of Heaven (Fig. S9A) as well as a famous US building: The White House (Fig. S9B). With electrical functions like controlled lighting in the drawing, the pattern could offer more electronic functions of the design. This electronics manufacture strategy would help incubate the new area of electronic architecture design. Meanwhile, it is also important for future educational training purpose.

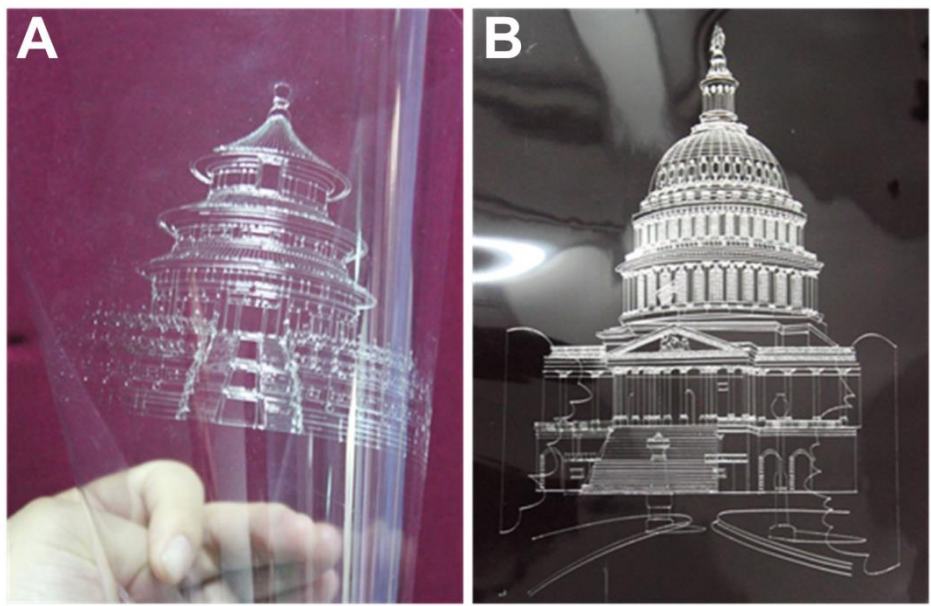

**Figure S9| Printed electronic building pattern.** (**A**) This printed building is called the Hall of Prayer for Good Harvests settling in Temple of Heaven as a principal building in Beijing, China. (**B**) Printed building of White House, Washington D. C., USA. (© Jing Liu)

A8. Printing complex structure and pattern of classical picture

　　With the big capability offered by the present technology, we were motived to print more complex electronic structures. One example can be found to print electronic picture of a famous Chinese painting: Along the River during the Qingming Festival. The original picture is rather large and not necessary to cover. For illustration purpose, we chose to print part of this drawing (see original in Fig. S10A). Before printing, it should be converted into a printable vector graph as shown in Fig. S10B. Then the final target pattern can be printed out. A result of such electronic picture is presented in Fig. S10C. For more illustration, presented in Fig. S10D is another printed electronics pattern for the famous building: The Great Wall. Clearly, with electrical elements inside the picture, a vivid presentation of ancient arts can thus be obtained. This bridges well the arts and science which may shed light for future combinational research.

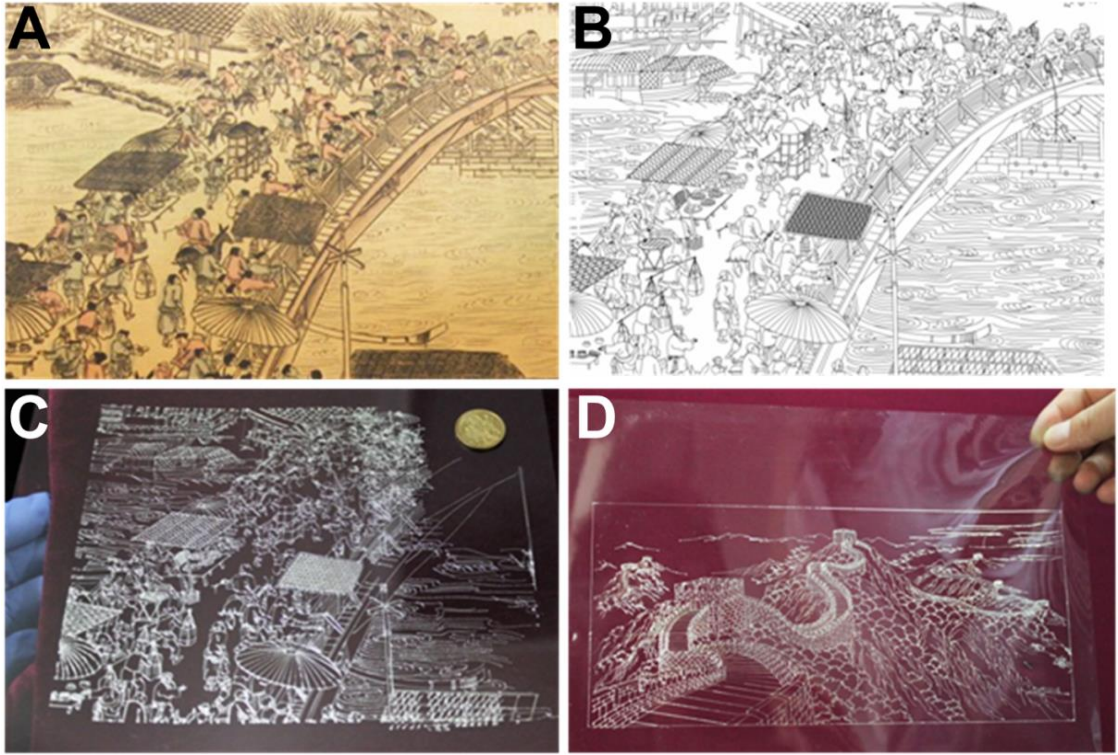

**Figure S10| The printed electronics of Chinese famous drawing.** (**A**) Original picture for Along the River During the Qingming Festival. (**B**) Vector graph for the original picture; (**C**) Printed metal drawing for Along the River During the Qingming Festival. (**D**) Printed metal drawing for China Great Wall. (© Jing Liu)

A9. Printing of conductive human portraits

Figure S11 illustrates the printed portraits of four scientists which are physicists Issac Newton (Fig. S11A), Albert Einstein (Fig. S11B), and two China scientist Xuan Wang (Fig. S11C) (inventor of Chinese Characters Phototypesetting which significantly innovate the printing technology) and Sheng Bi (Fig. S11D) (inventor of ancient China typograph).

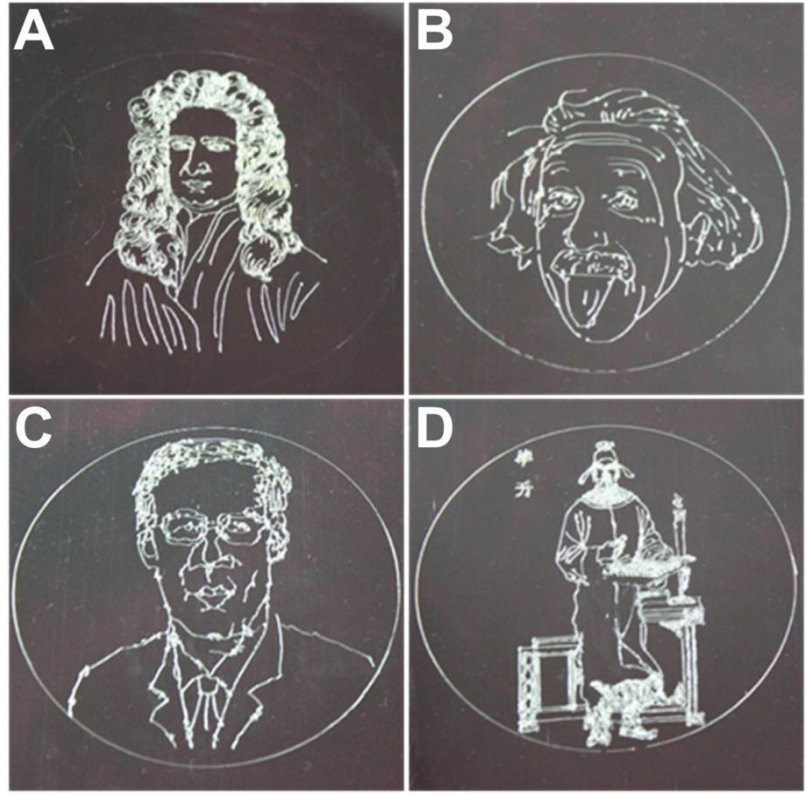

**Figure S11| Printed electronic human portrait.** (**A**) is a portrait of the great British scientist Isaac Newton. (**B**) is a portrait of Albert Einstein, who is one of the greatest scientists in 20th century. (**C**) is a portrait of Chinese scientist Xuan Wang, who is honored as the Father of Chinese Language Laser Typesetting. (**D**) is a portrait of an ancient scientist Sheng Bi in Song Dynasty China, who invented the world first known movable type printing press technology. (© Jing Liu)

A10. Printing of PCB board on flexible plastics

The most core function of the present method is to freely manufacture a printed-circuit-on-board (PCB) in a moment as desired. Figure S12 presented the design draft (Fig. S12A, D), printing process (Fig. S12B) and printed results of several PCB circuits (Fig. S12C, E). More discussion on such electronics can be found in the article. The present printed circuits are rather suitable for surface mounting electronic components. This significantly innovates the conventional PCB manufacturing technology.

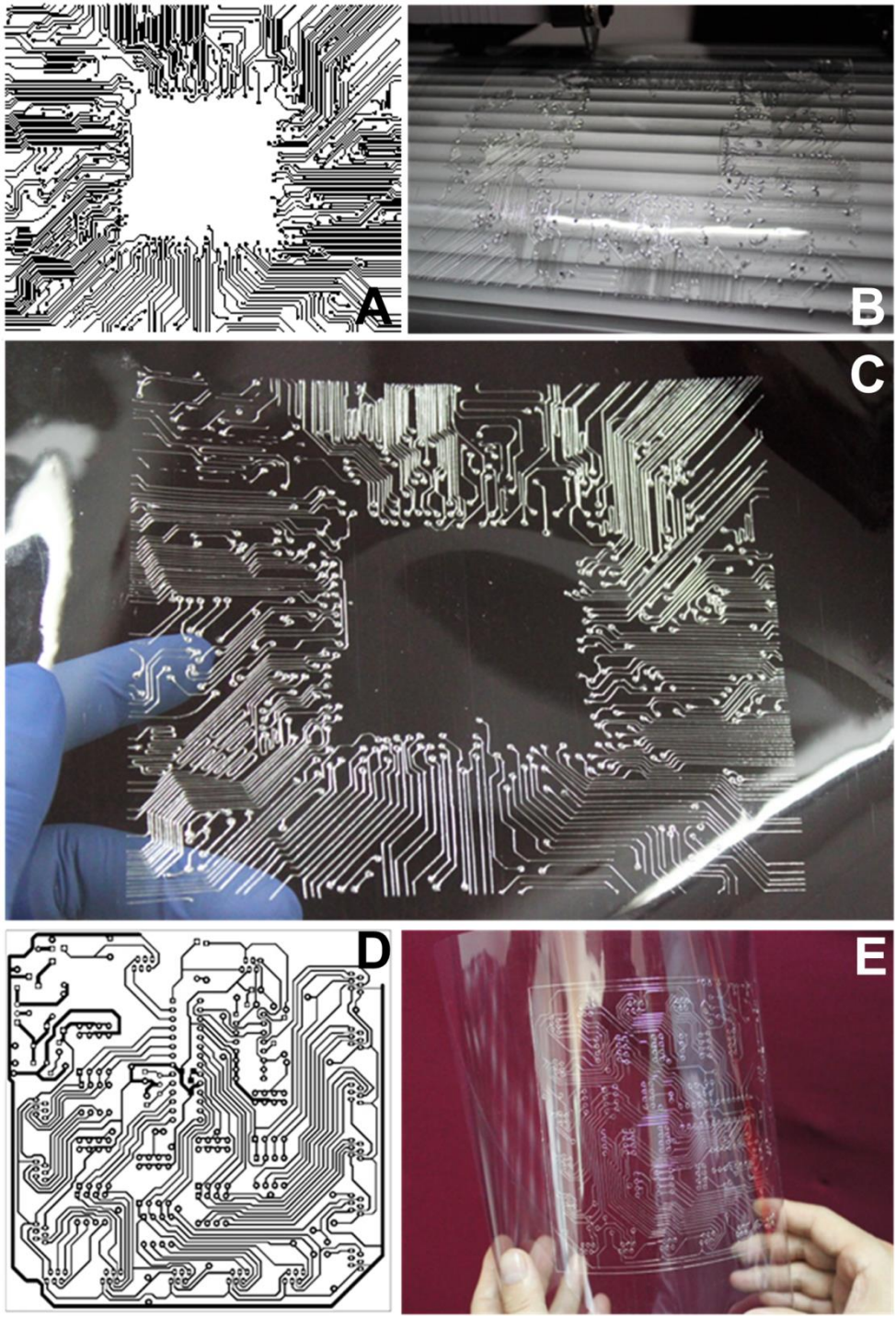

**Figure S12| Printed PCB in a few minutes using the liquid metal printer.** (**A**) Original PCB design draft; (**B**) PCB in printing on the machine; (**C**) Manufactured results of (A); (**D**) Additional PCB draft; (**E**) Printed PCB of (D). (© Jing Liu)

A11. Design and manufacture electronics as desired

As is increasingly realized, there is a huge demand to develop personal electronics manufacture way which was sometimes even classified as an important driving force for the coming "Third Industrial Revolution". The biggest barrier to impede the large scale practices of such endeavor lies in lack of cost effective machine itself and the limited capability of the ink. The invention of the present liquid metal printer makes the dream for electronics Do-It-Yourself a reality. Presented in Fig. S13 are just two of such typical working examples: electronic greeting card (Fig. S13A) and electronic decoration picture of the Great Wall (Fig. S13B). With lighting on, such drawings display more human's imaginations and are expected to play increasing roles in a wide variety of coming consumer oriented electronics.

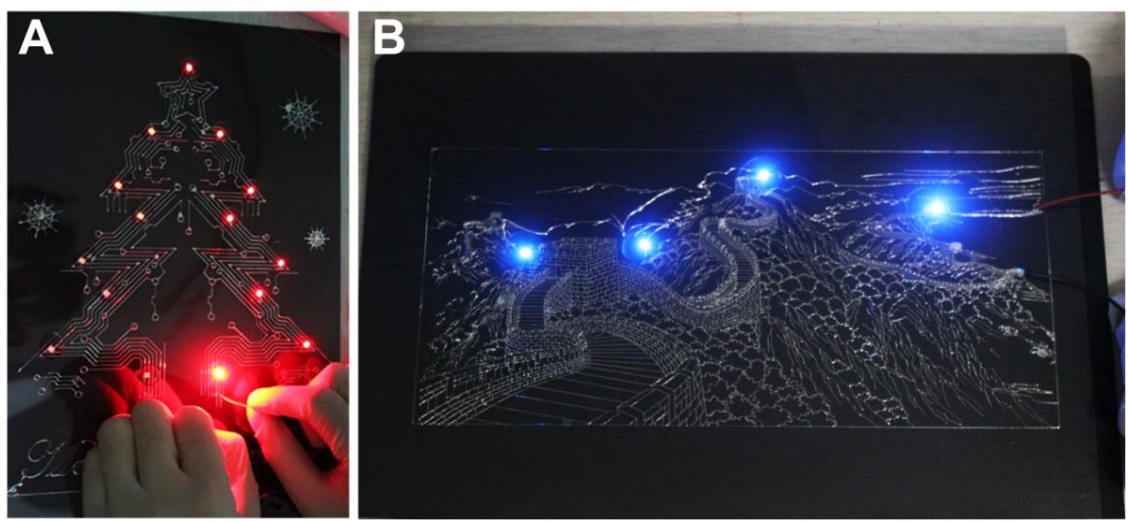

**Figure S13| Printed personal consumer electronics.** (**A**) Electronic greeting card with Christmas tree lighting on; (**B**) Electronic picture of China Great Wall with LED lighting on. (© Jing Liu)

A12. Package of printed electronics

If one wishes to completely ensure the environmental and mechanical stability of the printed electronic circuits, structures and patterns, a few commonly used materials such as PDMS or room temperature vulcanization (RTV) silicone rubber can be adopted to package the targets. Presented in Fig. 14 A and B are several such well packaged structures via RTV. The whole process in packaging each item just takes several minutes.

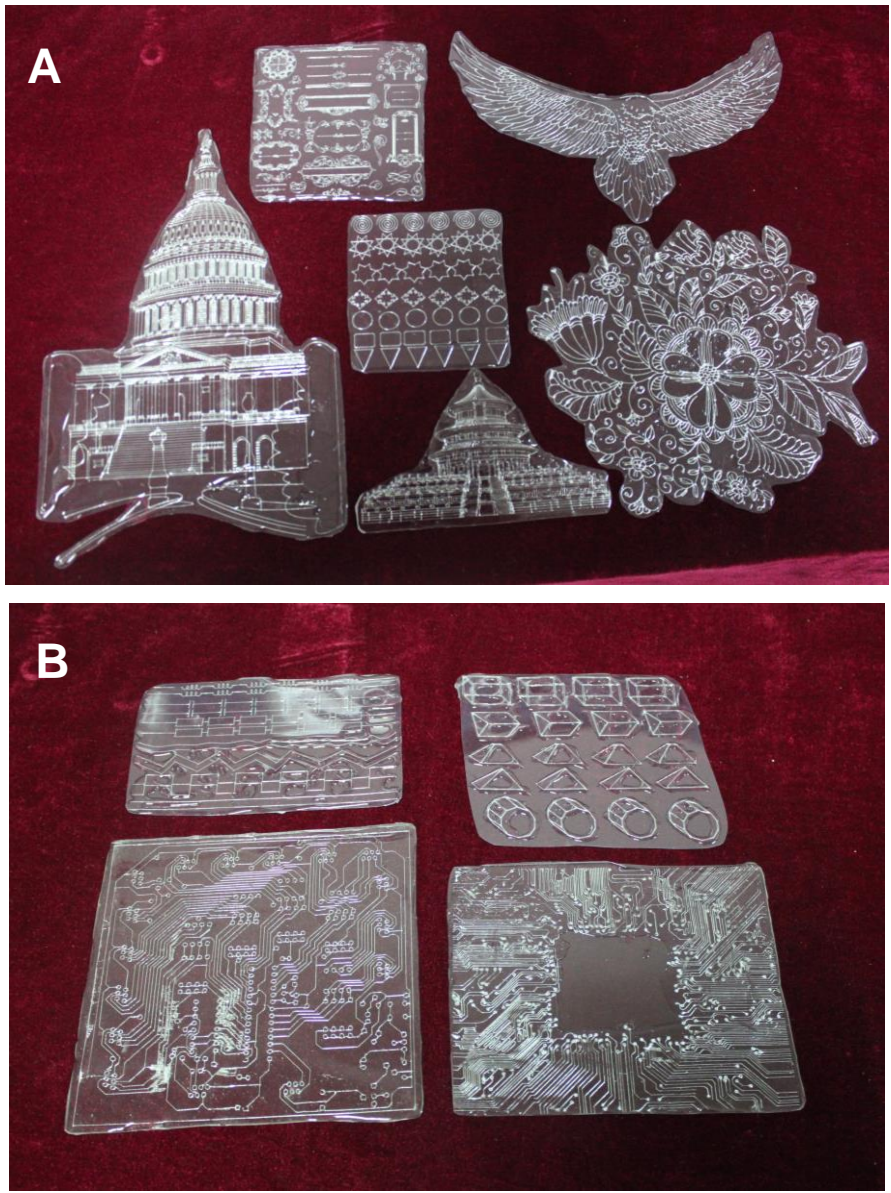

**Figure S14| Packaged printed electronic patterns.** (**A**) Printed electronic decoration arts. (**B**) Printed PCB or electronic elements. (© Jing Liu)

In conclusion, all these examples covering different areas comprehensively demonstrated the diverse capability of the liquid metal printer prototype in manufacturing various conductive structures and patterns spanning from a single line, curve to any desired complex paintings. This paved the way to directly and quickly fabricate flexible electronics as could as one can image. The total process looks just like printing a picture on the paper as it was often done in an office. The unique merit of the liquid metal printing technology lies in its entirely automatic controllability, extremely low cost, direct printing feature, and excellent adaptability. As is well known, conventional approaches of making a flexible circuit are generally complex, environment unfriendly, time and energy consuming, and thus expensive and hard to access by an ordinary user. The easy going feature, high-quality manufacture and extremely low cost merit of the present electronics printer make the goal of printing personal electronics at home a reality. It is expected to be widely used over the world in the coming time.

**Acknowledgements**

We would like to thank B. W. Chen for machine development and X. Xue for theoretical grid generation. Y.Z. performed the experiments, analyzed the data and wrote the manuscript; Z.Z.H. performed the theoretical simulation, analyzed the mechanism and wrote the manuscript; J.Y. performed part of the experiments, analyzed the data, and wrote the manuscript; J.L. conceived the project, designed the work, performed part of the experiments, and wrote the manuscript. All authors discussed the results and commented on the manuscript.